\documentstyle[preprint,aps,prd]{revtex}
\begin{document}
\title{Duality of massive gauge invariant theories in  arbitrary
space--time dimension\footnote{Accepted for publication in { \bf Phys.Rev.D}}}

\author{A.Smailagic \footnote{E-mail address: anais@etfos.hr} }
\address{Department of Physics, 
Faculty of Electrical  Engineering \\
University of Osijek, Croatia }
\author{E. Spallucci\footnote{E-mail address:spallucci@trieste.infn.it}}
\address{Dipartimento di Fisica Teorica\break
Universit\`a di Trieste,\\
INFN, Sezione di Trieste}
\maketitle

\bigskip

\begin{abstract}
 We show that dualization of  Stueckelberg--like massive gauge theories
 and $B\wedge F$ models,  follows form a general $p$--dualization of interacting
 theories in $d$ spacetime dimensions. This is achieved by a particular choice
 of the external current. 
 \end{abstract}

\newpage

Recently, various models of  $B\wedge F$ theories in $3+1$ dimensions,
either in relation to the Bosonization of fermionic models \cite{braz}, or
with respect to their dual Stueckelberg--like gauge models\cite{ind}, have been
considered. In this note we would like to show how to dualize $B\wedge F$ 
models in a general way. Also we show that such  dualization is a 
{\it special} case of a general formalism of  $p$--dualization of interacting 
field theories. Here we give a brief review of the general $p$--dualization 
procedure proposed in \cite{noi}.
As usual, one defines a parent Lagrangian from which original and dual theories
can be obtained. Parent Lagrangians are not uniquely defined and there exist
various methods for their constructions\cite{lind}.
 One of commonly used method is based on a  ``shift symmetry'' which has been
 also adopted in \cite{ind}. Instead
 we choose in our approach, what we believe to be, a simpler and more 
 transparent formulation of parent Lagrangian. It will
 describe dualization of {\it any interacting,} theory  of $p$--forms 
 in arbitrary spacetime dimension \cite{noi}. 
 Let us start with a theory of a potential $p$--form
$V(x)$ in arbitrary dimensions $d$ interacting with an {\it a priori}  external
current  $K(x)$. The dual  field $H$ is a $d-p-1$ form 
with the rank $p$ as $p\le d-1$. Parent Lagrangian is constructed as 

\begin{eqnarray}
L_P =&- &{1\over 2(d-p-1)!} H _{\mu_1\dots\mu_{d-p-1}}\, H 
^{\mu_1\dots\mu_{d-p-1}}
+{1\over (d-p-1)!} H^{\mu_1\dots\mu_{d-p-1}}\, 
F^*{}_{\mu_1\dots\mu_{d-p-1} }(\,V \,)
 \nonumber\\
  &+&{1\over (d-p-1)!} K^{\mu_1\dots\mu_{d-p-1}}  \, H_{\mu_1\dots\mu_{d-p-1}}
  \label{parent}
\end{eqnarray}

with the notation $F^*{}^{\mu_1\dots \mu_{d-p-1} }(\,V\, )= {1\over (p+1)!}
\epsilon^{\mu_1\dots \mu_{d-p-1} \mu_{d-p} \dots\mu_d}
F_{\mu_{d-p}\dots\mu_d}(V)$ and $F_{\mu_{d-p}\dots\mu_d}(\,V\, )\equiv 
 \partial_{\,[\,\mu_{d-p}}V_{\mu_{d-p+1}\mu_d\,]}$. Since, in general, dual
 field $H$ can be also a gauge potential  the external current $K$
 is not conserved, i.e. $\partial K\ne 0$. 

Dualization proceeds in the following way:\\
i) varying the parent Lagrangian  with respect to the field $H$ 
 leads to an  ``equation of motion'' 

\begin{equation}
\delta_H \, L_P=0\longrightarrow H^{\mu_1\dots\mu_{d-p-1}}= 
F^*{}^{ \mu_1\dots\mu_{d-p-1} }(\,V\, )
 + K^{\mu_1\dots\mu_{d-p-1}}  \label{bfk}
\end{equation}

The above equation,  inserted back into (\ref{parent}), gives  
the {\it interacting theory} for the  $V$  field described by the Lagrangian

\begin{eqnarray}
L_{V , K}&=& {1\over 2(p+1)!}\left[\, F^*{}_{\mu_1\dots\mu_{p+1}}(\,V\, )
+ K^{}_{\mu_1\dots\mu_{p+1}} \, \right]^2  \nonumber\\
 &=& -{1\over 2(p+1)!}\left[\, F_{\mu_1\dots\mu_{p+1}}(\,V\, )
-(-1)^{(p+1)(d-p-1)} K^*{}_{\mu_1\dots\mu_{p+1}} \, \right]^2
\label{la}
\end{eqnarray}

where, the $K^*$ is the Hodge dual of the current $K$, and the factor 
$(-1)^{(p+1)(d-p-1)}$ comes from epsilon contractions.    
One sees from (\ref{la}) that there is a current contact term in the theory
for $V$. Such terms are unavoidable products of dualization procedure, and it
turns out that they are necessary to make dualization consistent \cite{dj}, 
\cite{gris}. They will turn out to be essential for what we are going to prove.
\\
ii) The dual theory, on the other hand, is obtained varying parent Lagrangian 
with  respect to the field  $V$ which gives ``equation of motion'' as

\begin{equation}
\delta_V\, L=0\quad:\quad 
\partial^{\mu_1 } H^*{}_{\mu_1\dots\mu_{p+1}}=0
\label{bianchih}
\end{equation}

which has solution

\begin{equation}
 H_{\mu_1\dots\mu_{d-p-1}}=\partial_{\,[\,\mu_1}\, B_{\mu_2\dots
\mu_{d-p-1}\,]}\equiv H_{\mu_1\dots\mu_{d-p-1}}(B)
\label{bdu}
\end{equation}

Therefore, the dual field $H$ is defined through (\ref{bianchih}), and it
 turns out to be a field strength of the dual potential $B$.
(\ref{bdu}) inserted back in (\ref{parent}) gives 
the dual theory for the  field $B$, coupled to an external  current $K$ 
 
\begin{eqnarray}
L_{B,K}= &-&{1\over 2(d-p-1)!}\,  H_{\mu_1\dots\mu_{d-p-1}}(B)\, 
H^{\mu_1\dots\mu_{d-p-1}}(B) + \nonumber\\
 &+& {1\over (d-p-1)!}K^{\mu_1\dots\mu_{d-p-1}} \,
H_{\mu_1\dots\mu_{d-p-1}}(B)
\label{lb}
\end{eqnarray}

The end result of our dualization procedure is that {\it to any interacting
theory} of the $p$--form $V$, described by the  Lagrangian  (\ref{la}), 
corresponds  a {\it  dual interacting theory} in terms of the dual potential 
$B$, described by the Lagrangian  (\ref{lb}). One can verify that this procedure
reproduces all known dual theories as described for example in \cite{lind}
( see final discussion ).\\
At this point one may wonder how can this procedure produce
dualization of $B\wedge F$ theories? Before we address the above question
we rewrite, by partial integration, the interaction term of the dual theory 
as 

\begin{eqnarray}
L_{B, int.} & \equiv &
 {1\over (d-p-1)!}K^{\mu_1\dots\mu_{d-p-1}} \, H_{\mu_1\dots\mu_{d-p-1}}(B)
 \nonumber\\
 & = &
 -{1\over(d-p-2 )!}\left(\partial_{\mu_1}K^{\mu_1\mu_2\dots\mu_{d-p-1}}\right)
 B_{\mu_2\dots\mu_{d-p-1}}
\label{lbb}
\end{eqnarray}

Also notice that the described dualization procedure will not be affected
if one adds to the parent Lagrangian (\ref{parent}) any function of the
external current $K$. Therefore we choose to modify (\ref{parent}) , for
the reason to become clear soon, as

\begin{equation}
L_P\longrightarrow L_P +{1\over 2m^2(d-p-2)!}\left(\,
  \partial_{\mu_1} K^{\mu_1\mu_2\dots\mu_{d-p-1}}\,\right)^2  
\end{equation}

and a mass parameter $m$ has to be introduced for dimensional reasons.\\
 What we claim is that the general interaction term (\ref{lbb})
 is equivalent to the $B\wedge F$ terms in special cases. To see this, let us 
 rewrite the external current $K$ as the Hodge dual of a {\it new } gauge field
 $A$.  By ``new gauge field'' we mean independent of the gauge fields $V$ and
  $H$, which are subject to the dualization procedure.

\begin{equation}
K^{\mu_1\dots\mu_{d-p-1}}=
{m\over (p+1)! }
\epsilon^{\mu_1\dots\mu_{d-p-1}\mu_{d-p}\dots\mu_d}\, A_{\mu_{d-p}\dots\mu_d}
\label{KA}
\end{equation}

Inserting (\ref{KA}) into (\ref{lbb}) we find that the interaction term 
$H(B)K$ translates into

\begin{equation}
L_{B,int.}\equiv   {m\over d! } \epsilon^{\mu_1\dots\mu_d}
B_{\mu_1\dots\mu_{d-p-2}}\partial_{\,[\,\mu_{d-p-1}}
A_{\mu_{d-p}\dots\mu_d\,]}\equiv m\, B^{\mu_1\dots\mu_{d-p-2}}F^*{}_
{\mu_1\dots\mu_{d-p-2}}(A)
\end{equation}

which is a {\it generalized $B\wedge F$ term.}  ${\cal QED}$. Then,
the Lagrangian (\ref{la}) of the original theory for $V$ becomes 

\begin{equation}
L_{V, A}= -{1\over 2(p+1)!}\, \left(\,\partial_{\,[\,\mu_1}
V_{\mu_2\dots\mu_{p+1}\,]} + m\,A_{\mu_1\dots\mu_{p+1}} \, \right)^2  
-{1\over 2(p+2)! }F^2{}_{\mu_1\dots\mu_{p+2}}(A)
\label{lstuck}
\end{equation}

(\ref{lstuck}) makes clear the choice of the function $f(K)$, previously 
introduced in the parent Lagrangian. It gives a kinetic term of the 
new gauge field $A$. Therefore, starting from a general interacting theory,
we obtain gauge invariant, Stueckelberg--like Lagrangian, 
of a {\it massive} $p+1$--form  potential $A$,
where the starting $p$--form $V$ field acts as a Stueckelberg conpensator. \\
The dual $B\wedge F$ theory of such  Stueckelberg--like model follows directly
from (\ref{lb}) as

\begin{eqnarray}
L_{BF}= &-&{1\over 2(d-p-1)!}\, H^2{}_{\mu_1\dots\mu_{d-p-1}}(B)  + 
{1\over d!}  \epsilon^{\mu_1\dots\mu_d}  B_{\mu_1\dots\mu_{d-p-2}}
\partial_{\,[\,\mu_{d-p-1}}A_{\mu_{d-p}\dots\mu_d\,]}+\nonumber\\
&-&{1\over 2(p+2)! }\, F^2_{\mu_1\dots\mu_{p+2}}(A)
\label{lbd}
\end{eqnarray}

This is the general result we were set to prove in this note.\\
To make the final result more transparent let us look at specific case of 
$(3+1)$ dimensions. We find the following models:\\
\begin{enumerate}
\item choice of rank $p=0$ implies identification of various fields as 
 $V\rightarrow\Phi$, $B\rightarrow B_{\mu\nu}$, 
$A\rightarrow A_\mu$. From the above formulae one finds  the Stueckelberg 
gauge theory of the massive vector field as
 
\begin{equation}
L_\Phi= -{1\over 2}\left(\, \partial_\mu\Phi + m\, A_\mu\,\right)^2-{1\over
4}\, F^2{}_{\mu\nu}(A)
\label{fi4}
\end{equation}

and its dual $B\wedge F$ theory, in terms of the Kalb--Ramond tensor field, 
which is given by the Lagrangian

\begin{equation}
L_B=-{1\over 2\cdot 3!}\, H^2{}_{\mu\nu\rho}(B)  + 
{m\over 4!}\, \epsilon^{\mu\nu\rho\sigma} \, B_{\mu\nu}\partial_{\,[\,\rho }\,
A_{\sigma\,]}-{1\over 4}F^2{}_{\mu\nu}(A)
\label{lb4}
\end{equation}

\item choice of rank $p=1$ implies identification of various fields as 
 $V\rightarrow\Phi_\mu$, $B\rightarrow B_{\mu}$, $A\rightarrow A_{\mu\nu}$.
Stueckelberg--like formulation of the massive, gauge invariant, Kalb--Ramond
model  is given by

\begin{equation}
L_A= -{1\over 4}\left(\,\partial_{[\,\mu}\Phi_{\nu\,]} + m
A_{\mu\nu} \, \right)^2  
-{1\over 2\cdot 3! }F^2{}_{\mu\nu\rho}(A)
\end{equation}

while the dual $B\wedge F$ theory is described by the Lagrangian 
\begin{equation}
L_B=-{1\over 4} H^2{}_ { \mu\nu } (B)  + 
{m\over 4!} \epsilon^{\mu\nu\rho\omega}\, B_\mu\,
\partial_{\,[\,\nu}A_{\rho\sigma\,]}-{1\over 2\cdot  3!}F^2{}_{\mu\nu\rho}(A)
\label{lbd1}
\end{equation}

\item
Last possible choice is $p=2$, which implies 
$V\rightarrow \Phi_{\mu\nu}$, $B\rightarrow \phi$, 
$A\rightarrow A_{\mu\nu\rho}$ leading to  Stueckelberg Lagrangian 

\begin{equation}
L_A= -{1\over 2\cdot 3!}\left(\,\partial_{[\,\mu}\Phi_{\nu\rho\,]} + m
A_{\mu\nu\rho} \, \right)^2  
-{1\over 2\cdot 4! }F^2{}_{\mu\nu\rho\sigma}(A)
\end{equation}

while the dual $B\wedge F$ theory is described by the Lagrangian 
\begin{equation}
L_B=-{1\over 2}\left(\, \partial_\mu\phi \,\right)^2 + 
{m\over 4!} \epsilon^{\mu\nu\rho\sigma}\, \phi\,
\partial_{\,[\,\mu}A_{\nu\rho\sigma\,]}-{1\over 2\cdot4!}
F^2{}_{\mu\nu\rho\sigma}(A)
\label{lbd2}
\end{equation}
and so on {\it ad nauseam,} for other spacetime dimension.
\end{enumerate}
The first and the last models  in the above  examples are exactly those
described in \cite{ind} while we showed that the general formalism gives also
a third model based on vector--to--vector  gauge field dualization. 
As a conclusion, we have shown how  Stueckelberg--like
massive gauge theories of an external gauge potential $A$, introduced as
the Hodge dual of the external current $K$, and their dual $B\wedge F$ models,
represent special cases of a general dualization procedure of {\it interacting}
theories. Realization of  Stueckelberg--like model is due
to the fact that dualization procedure combines the field strength of the
potential $V$ and the external current $K$  in the form (\ref{la}), thanks to
the existence of the contact term $K^2$.
  Our general procedure thus enables to consider dual theories at will
  by appropriate choice of the external current. Note that the choice
  $A=0$ in the Stueckelberg--like four dimensional examples 
  reproduces all known four dimensional dual free theories, i.e. scalar to
  Kalb--Ramond (either way), and vector to vector. This proves the power and
  generality of the proposed method.

	\end{document}